\documentclass{JHEP3}
\newcommand{\lsim}{\lesssim}
\newcommand{\gsim}{\gtrsim}
\usepackage{epsfig}

\title{Reopening the window on charged dark matter}
\author{Leonid Chuzhoy \\ Department of Astronomy and Astrophysics, The University of Chicago, 5640 S. Ellis, Chicago, IL 60637, USA; chuzhoy@oddjob.uchicago.edu}
\author{Edward W. Kolb \\ Department of Astronomy and Astrophysics, Enrico Fermi Institute, and Kavli Institute for Cosmological Physics, The University of Chicago, 5640 S. Ellis, Chicago, IL 60637, USA}

\abstract{We reexamine the limits on charged dark matter particles.
We show that if their mass and charge fall in the range
$100(q_X/e)^2\lsim m_X\lsim10^{8}(q_X/e)$ TeV, then magnetic fields
prevent particles in the halo from entering the galactic disk, while
those initially trapped inside are accelerated through the Fermi
mechanism and ejected within about $0.1-1$ Gyrs. Consequently,
previous constraints on charged dark matter based on terrestrial
non-observation are invalid within that range. Further, we find that
charged massive particles may simultaneously solve several
long-standing astrophysical problems, including the underabundance
of dwarf galaxies, the shallow density profiles in the cores of the
dwarf galaxies, the absence of cooling flows in the cores of galaxy
clusters, and several others.}


\begin{document}

\section{Introduction}
That most of the mass in the Universe is comprised of unknown
non-baryonic particles has been in the mainstream of cosmology for
almost a generation. Yet attempts to detect these particles only
succeeded in excluding many attractive candidates. Recently,
underground detectors have also produced strong constraints on the
weakly interactive massive particles (WIMPs), limiting their
interaction cross-section with nucleons to about  $10^{-43}\;{\rm
cm^2}$ \cite{XENON,CDMS}.  However, in this paper we show that the
reason for non-detection might not be the low interaction strength,
but rather the depletion of dark matter from the galactic disk. This
could occur if dark matter particles were charged and their
charge-to-mass ratio fell into a specific range. Furthermore, we
find that charged massive particles (CHAMPs) may simultaneously
solve several long-standing astrophysical problems, including the
underabundance of dwarf galaxies (known as the missing satellite
problem), the too shallow density profiles in the cores of the dwarf
galaxies, the absence of cooling flows in the cores of galaxy
clusters, and several others \cite{Mo,Kl,NS,SB,cool1,cool2,cool3}.

It was proposed long ago that a stable particle with a unit charge can act as
nearly collisionless dark matter if its mass is sufficiently high \cite{DeR}.
However, subsequent experiments using both terrestrial and satellite data
failed to detect any traces of such particles in our vicinity (see e.g.,
Dimopoulos et al. 1990).  The less appealing option of dark matter particles
with fractional charge \cite{Hol} is also subject to strong observational
constraints \cite{DHR}. However, as shown by our calculations, if the charge
and mass of CHAMPs fall in the range $100(q_X/e)^2\lsim m_X \lsim
10^{8}(q_X/e)$ TeV, then their absence in the galactic disk (and hence
non-detection on Earth) can be naturally explained by their interaction with
galactic magnetic fields. In brief, galactic magnetic fields, which are
observed to be parallel to the disk, prevent halo CHAMPs from penetrating the
disk. \footnote{A direct evidence of efficient segregation of charged particles by galactic magnetic fields comes from the cosmic rays, whose typical life-time inside the disk is of order $10^7$ years, i.e., around $10^5$ times longer than would be required for their direct escape.} At the same time, interactions with expanding supernovae (SN) remnants boost the kinetic energy of CHAMPs initially present inside the disk, leading to their expulsion on a time-scale of 0.1-1 Gyrs. This mechanism is similar to the one that produces cosmic rays, however, it is much more efficient due to the largely dissipationless nature of the CHAMPs.


The paper is organized as follows:  In \S 2 we consider the chemistry of
CHAMPs, baryons, and electrons. In \S 3 we describe the mechanism for CHAMP
ejection from the Milky Way disk. In \S 4 we describe the changes CHAMPs would
introduce into galaxy formation process. We summarize our results in \S 5.

\section{ CHAMP chemistry }
CHAMPs may recombine with baryons and electrons, which would affect their
charge state, and hence their interactions with electromagnetic fields and
their mean-free-path. Depending on the initial charge, we can group
CHAMPs into three categories.

{\it I:} The chemical properties of CHAMPs with positive unit charge would
be very similar to protons. Consequently, in the interstellar medium the
fraction of free CHAMPs ($X^{+}$) and CHAMPs recombined with electrons
($X^{+}e$) would mirror the local ionization state of hydrogen. Likewise,
CHAMPs with integer charge greater than unity would mirror the behavior of
heavier elements.

{\it II:} Following nucleosynthesis, CHAMPs with negative unit charge can
recombine with baryons, forming neutral or positively charged particles
($X^-p$ or $X^-\alpha$). Since the binding energy increases with the mass and
charge of the baryon, at later epochs some of the CHAMPs can undergo
charge-exchange reactions, such as $X^-p+\alpha\rightarrow X^-\alpha+p$ or
$X^-p+Li\rightarrow X^-Li+p$. Unfortunately, these reaction rates are very
uncertain, so at present most of these particles could be either free, bound
to the helium ions, or bound to protons \cite{Dal}.

{\it III:} CHAMPs with fractional charge might also recombine with ordinary
matter, though with smaller binding energies the combinations would be more
vulnerable to dissociation. Unlike CHAMPs with integer charge, CHAMPs with
fractional charge can never form neutral particles. Hence, unless their charge
to mass ratio is negligible they will be unable to masquerade as neutrals.

It seems most plausible to assume that charged dark matter would be made up by
the equally numerous particles and anti-particles with a unit charge ($X^{-}$
and $X^{+}$). In this case, in the interstellar medium, where hydrogen is
mostly ionized, a larger fraction of CHAMPs would form charged particles (e.g.,
$X^{+}$ and $X^{-}\alpha$), while a smaller fraction would masquerade as
neutrals (e.g., $X^{-}p$), which would be unaffected by magnetic fields and
have a low scattering cross section.  Alternatively, if CHAMPs have fractional
or only positive charge (the latter possibility can not be excluded, given the
example of the baryon asymmetry), then charged particles can make up virtually
all of the dark matter.

\section{Dark matter in the galactic disk}
The large-scale magnetic field in the Milky Way, with strength of order
1-10 $\mu G$, is mostly parallel to the plane of the Galactic disk.
Consequently, the particles in the halo are unable to penetrate the disk and
those inside it to escape, unless their gyroradius,
\begin{equation}
\label{gyro}
R_g=10^{-9}{\rm pc}\left(\frac{m_X}{m_p}\right)
\left(\frac{e}{q_X}\right)\left(\frac{v_X}{300\;{\rm km\ s}^{-1}}\right)
\left (\frac{B}{1{\rm \mu G}}\right)^{-1},
\end{equation}
is larger than the typical height of the disk (about $100$ pc). In Eq.\
(\ref{gyro}), $m_p$ is the proton mass, $v_X$, $q_X$, and $m_X$ are the
CHAMP's velocity, charge and mass.  With the velocity dispersion in the halo
of order $300$ km s$^{-1}$, crossing the magnetic lines requires the CHAMP
mass to be above about $10^{8}(q_X/e)$ TeV. (Inside the disk the energy of
charged particles can increase, which may allow even lower mass particles to
escape.)

The main processes affecting the energy of charged particles in the disk are
interactions with dynamic magnetic fields driven by supernovae (SN) explosions
(the Fermi mechanism), radiative cooling, and Coulomb scatterings, which
respectively inject, dissipate, and redistribute energy in plasma. The action
of the Fermi mechanism can be described as following: A charged particle
traveling in the interstellar medium would repeatedly scatter off the magnetic
field lines of expanding SN remnants. In each scattering the particle velocity increases, thereby decreasing the average time between scatterings. In the
absence of other processes, this leads to an exponential momentum growth.

The energy injection rate from SN remnants into each particle species is
roughly proportional to their thermal pressure, so the time-scale for momentum
increase, $\tau_\textrm{acc}=v/(dv/dt)$ is also roughly the same for all
charged species. Since injecting more energy into baryons than can be
dissipated by radiative cooling would lead to a suppression of star formation,
$\tau_\textrm{acc}$ must generally be close to $\tau_\textrm{cool}$, about
$10$ Myrs in the present-day Milky Way disk \cite{McO}.

While baryons and CHAMPs receive roughly the same energy per unit mass from
SNs, radiative cooling is limited to the former. Thus, energy balance
may be preserved only by frequent Coulomb scatterings.  For particles whose
velocities are smaller than thermal velocities of electrons, scatterings with
protons are dominant, with a relaxation time-scale of
\begin{eqnarray}
\label{txp}
\tau_{\textrm{rel},p} & \approx & 300\; \textrm{years} \;
\left(\frac{e}{q_X}\right)^{2}
\left(\frac{m_X}{m_p}\right)
\left(\frac{v_X}{100\;{\rm km\ s}^{-1}}\right)^3 \nonumber \\
& & \times \left(\frac{n_p}{\rm 10^{-2} cm^{-3}}\right)^{-1},
\end{eqnarray}
where $n_p$ is the proton number density. Those with higher velocities lose
energy mainly to electrons, with a relaxation time-scale of
\begin{eqnarray}
\label{txe}
\tau_{\textrm{rel},e} & \approx & 200\; \textrm{years} \;
\left(\frac{e}{q_X}\right)^{2}
\left(\frac{m_X}{m_p}\right)
\left(\frac{v_X}{1000\;{\rm km\ s}^{-1}}\right)^3 \nonumber \\
& & \times \left(\frac{n_p}{\rm 10^{-2} cm^{-3}}\right)^{-1}.
\end{eqnarray}
If for CHAMPs, $\tau_\textrm{rel}$ is larger than $\tau_\textrm{acc}$, the
CHAMPs velocities would grow until they can no longer be contained in the disk.
The escape can follow one of two routes:  At early stages CHAMPs are likely to
comprise a large fraction of the disk mass. If
$\tau_\textrm{acc}<\tau_\textrm{rel}$, eventually their kinetic energy becomes
larger than the total (CHAMPs plus baryons) binding energy of the disk and they
escape from the disk, sweeping with them the magnetic field lines and the
ionized gas which is tied to the magnetic field. CHAMP escape, therefore, would
be accompanied by a significant depletion of baryons from the disk, as only
neutral gas clouds may stay behind.  At a later epoch, when the CHAMP abundance
is too low to sweep baryons with them, their energy can continue to grow until it is high enough for crossing the disk magnetic field lines.

In both cases, in order to satisfy the condition for escape,
$\tau_\textrm{acc}<\tau_\textrm{rel}$, CHAMP velocities must exceed a
critical threshold \footnote{We assume conservatively that CHAMPs can escape
only when their velocity is higher than the typical electron thermal velocity,
i.e., when $\tau_\textrm{rel} \approx \tau_{\textrm{rel},e}\ll
\tau_{\textrm{rel},p}$.}
\begin{eqnarray}
\label{vcrit} v_{\rm crit} & \sim & 4 \times 10^4 \; \textrm{km s}^{-1}\;
\left(\frac{m_X}{m_p}\right)^{-1/3}
\left(\frac{q_X}{e}\right)^{2/3} \nonumber \\
& & \times \left(\frac{n_p}{10^{-2} \textrm{cm}^{-3}}\right)^{1/3}
 \left( \frac{\tau_\textrm{acc}}{10\; \textrm{Myr}} \right)^{1/3}.
\end{eqnarray}
Beyond this threshold a particle would on average gain more energy from SNs
than lose in the Coulomb scatterings, so most CHAMPs would have to cross it
only once in order to escape.

SN produced shockwaves with velocities comparable to the thermal velocities of
protons in the hot ionized medium (about $100$ km s$^{-1}$) are expected to
travel across most of the galactic disk every $\tau_\textrm{cool}\sim 10$ Myr.
Energy conservation in a point explosion requires that the probability of a
particle being hit by a shock with velocity greater than $V_s$ scales as
$V_s^{-2}$, so that shocks with $V_s\gsim 1000$ km s$^{-1}$ are expected to
pass about once per Gyr. From Eq.\ (\ref{vcrit}) we see that at the
present-day conditions of the Milky Way disk $V_s\sim 1000$ km s$^{-1}$ would
be sufficient to launch the exponential acceleration of CHAMPS with mass
$m_X\gsim 100 (q_X/e)^2$ TeV, which in about $100$ Myrs would lead to their
escape from the disk.  Since the SN activity seems to be much higher in past
and $\tau_\textrm{acc}$ respectively shorter, this is likely to be a
conservative estimate.

Thus, for masses in the range $100(q_X/e)^2\lsim m_X\lsim 10^{8}(q_X/e)$ TeV,
CHAMPs in the disk should have been severely depleted by SN shock
acceleration, and CHAMPs in the halo will be unable to penetrate into the disk
because of interactions with the magnetic field \footnote{The fact that SN shocks could make a strong impact on CHAMPs has already been noted by Dimopoulos et al. (1990). However, in their scenario they did not consider the segregation of CHAMPs from the disk by magnetic fields. Consequently, they expected that low mass CHAMPs would overheat the baryonic disk, leading to its destruction.}.

\section{Observational signatures of CHAMPs}
The interaction of CHAMPs with ordinary matter depends on its charge to mass
ratio, $q_X/m_X$. Observations put severe limits on the abundance of CHAMPs
with $m_X\lsim 100 (q_X/e_X)^2$ or $m_X\gsim 10^{8} (q_X/e_X)$ TeV that are
not expelled from the Milky Way disk, and whose potential impact is therefore
negligible. By contrast, CHAMPs in the range  $100 (q_X/e_X)^2 \lsim m_X\lsim
10^{8} (q_X/e_X)$ TeV can strongly affect the visible universe by interacting
with ordinary matter via the mediation of magnetic fields, and, for those
belonging to the lower part of this range, $100 (q_X/e_X)^2 \lsim m_X \lsim
1000 (q_X/e_X)^2$ TeV, directly by  Coulomb scatterings.

\subsection{The Effects of Magnetic interaction}
\subsubsection{Dark matter density profiles}
Unlike CDM halos, which remain almost static after the end of the accretion
phase, CHAMP halos can undergo drastic changes following the formation of
stars. Unless the CHAMP's charge is large enough for the Coulomb scatterings
to be important (the possibility discussed in the next section), the energy
injection from SNs will continuously reduce the binding energy of the
dark-matter halo. In large elliptical galaxies, with large binding energies
and low luminosity per unit mass, the effect on the dark-matter density
profiles would be moderate. In spiral galaxies like the Milky Way, the total
mechanical input from the SNs is comparable to the binding energy. However,
segregating CHAMPs from the stellar disk should greatly decrease the
fraction of energy going into the dark matter, so the effect is again likely
to be moderate. By contrast, in spherical dwarf galaxies the energy
injection from SNs may drastically reduce the steepness of the dark-matter
density profile, possibly leading to the expulsion of the CHAMPs from the
galaxy.

This mechanism might explain the absence in dwarf galaxies of a central
density cusp, which is predicted by the numerical simulations.

\subsubsection{Angular momentum of the galactic disk}
Scatterings of the halo CHAMPs from the magnetic field lines of the disk
create momentum exchange between the disk and the halo. Since the dark-matter
mass density close to the disk surface is estimated to be comparable to the
baryon density, the time-scale for momentum equilibration is roughly given by
\begin{equation}
\label{momeq}
\tau_\textrm{ang}=\frac{R_h}{V_X}\approx 5\times 10^5\; \textrm{years},
\end{equation}
where $R_h\approx 100$ pc is the typical disk height and $V_X\approx 200$ km
s$^{-1}$ is the tangential velocity of the CHAMPs. Since $\tau_\textrm{ang}$
is much shorter than the time-scale for momentum exchange via dynamical
friction, we suspect that the history of the disk formation may be very
different than in the standard model. In particular, it would be interesting
to see whether numerical simulations incorporating this effect would be able
to solve the so called angular momentum problem (i.e., the discrepancy between
the low disk momentum produced by current simulations and the much higher
observed  momentum).

\subsubsection{Break-off of hydrostatic equilibrium}
Following the collapse into a halo, the gas and the CHAMPs gain the same
amount of energy per unit mass, which should result in similar distribution
profiles. However, non-gravitational processes, such as cooling, affect
each component differently, leading to different dynamics. Unlike the CDM
model where gas is free to settle at its new equilibrium distribution, the
coupling to the dark matter via the magnetic field can support the gas against
gravity when its own pressure is insufficient.

This process may explain the discrepancy between the mass estimates in galaxy
clusters made independently from lensing and X-ray observations. The latter,
which rely on the assumption of hydrostatic equilibrium, is in agreement with
lensing observations at large radii where the cooling exceeds the Hubble time.
However, at the central region the lensing often overpredicts the mass by a
factor of a few \cite{Allen}.

\subsection{Effects of collisional interactions}
If the CHAMP mass is significantly above the proton mass, then following their
accretion by the halo the CHAMPs would typically have much higher kinetic
energies than baryons. Consequently, collisions of CHAMPs and baryons result
in a net heat inflow for the gas, which can prevent its accretion unless
balanced by strong cooling.

To illustrate this effect we calculate the critical temperature, $T_{cr}$, at
which heating by CHAMPs is balanced by radiative cooling in halos of different
size.  When $T_{cr}$ is above the virial temperature, $T_{vir}$, gas is unable
to collapse into the halo because of thermal pressure. Conversely, when
$T_{cr}< T_{vir}$, the gas density will continue to rise, which can further
increase the cooling rate.

Figure 1 shows the critical temperature in halos with different CHAMP velocity
dispersions. For this example we assumed that CHAMPs have a mass of $m_X=400$
TeV and a charge +1 (i.e., most is made of $X^+$ or $X^-\alpha$). The
ionization fraction of CHAMPs should be very similar to that of hydrogen. We
expect the scattering cross-section of CHAMPs that recombined with electrons
with hydrogen atoms to be comparable to that of hydrogen atoms, $10^{15}
{\rm cm^2}$.\footnote{It is possible that the near-degeneracy of the electron
energy levels in the CHAMP and the hydrogen atom can boost the cross-section
by one or two orders of magnitude, but in our case this would not make a
significant change.}  We consider separately the epochs prior to hydrogen and
helium reionization, respectively, when collisional excitation of HI and HeII
atoms were the dominant cooling processes, and the epoch when the helium is
fully ionized and the metal-poor gas can cool only by Bremsstrahlung.  The
comparison of the critical and the virial temperatures shows that at different
epochs heating by CHAMPs suppresses gas accretion in halos with velocity
dispersion below about 10, 20 and 40 km s$^{-1}$, respectively. This would
change  the reionization history of the Universe compared to standard model
since at early times because heating by CHAMPs suppresses the formation of
minihalos which would shift the beginning of star formation to a later epoch,
when halos with $V_X\gsim 10$ km s$^{-1}$ start to form. In the latter case,
by contrast, heating by CHAMPs is likely to boost the formation of radiation
sources. Heating by CHAMPs would significantly increase the Jeans mass, thus
biasing the first sources towards larger masses and emissivity. Since cooling
in the metal-poor gas is very inefficient below $10^4$K, it is further
conceivable that these halos would not be able to form stars at all,
collapsing into supermassive black holes instead. This would explain the
appearance of supermassive black holes already at $z>6$, which would be
problematic if the black holes grew by accretion starting with stellar mass
seeds \cite{TH}.

As reionization of hydrogen progresses, first generation halos stop
forming, thus slowing the production of ionizing photons, until new halos with
$V_X\gsim 20$ km s$^{-1}$ start to form in large numbers. After the end of
helium reionization only galaxies with $V_X\gsim 40$ km s$^{-1}$ may form,
with smaller halos remaining dark.

The above reionization scenario offers answers to several existing problems:
The end of hydrogen reionization is observed to be at $z\sim 6$, while most of
hydrogen must be ionized already at $z\gsim 11$ \cite{Sp}; present
star-formation rates are insufficient to produce early reionization \cite{Gn};
and the number of observed galaxies with rotational velocities below 40 km
s$^{-1}$ is much lower than expected \cite{Mo,Kl}.

\begin{figure}
\centering
\includegraphics[width=0.8\textwidth]{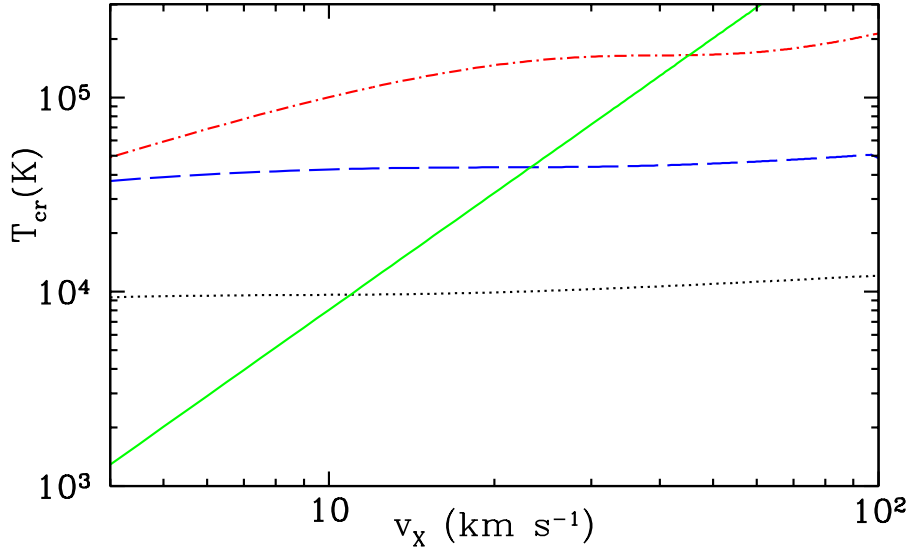}
\caption{The critical temperature
vs. the CHAMPs velocity dispersion. The dotted and dashed lines
correspond to the epochs before HI and HeII reionization,
respectively. The dot-dashed line corresponds to the epochs after
HeII  reionization. The solid line shows the virial temperature,
$T_{vir}\sim 2m_pV_X^2/3k_B$. \label{fig:Tcrf}}
\end{figure}

\subsection{Dark matter annihilation}
If the dark matter is composed of particles and anti-particles, their
annihilation signature may be observable. The evolution of CHAMPs annihilation rate has
several peculiar features that potentially can make its signature distinct from neutral dark matter particles.

The attractive Coulomb potential between free $X^+$ and $X^-$, increases the
annihilation cross-section by a factor $\sim c/v$ (the Sommerfeld-Sakharov correction). Hence,
after the CHAMPs become non-relativistic, their annihilation rate
falls at a slower rate than for the CDM model. By contrast,
other models with Sommerfeld enhancement, where the effect is achieved by self-interaction of dark matter particles,
produce an even slower decline of the annihilation rate. Since they are not collisionally coupled to radiation, their
kinetic energies fall as $(1+z)^2$, rather than $(1+z)$ as in the CHAMPs case.

Around $z\sim 10^6$ the CHAMPs annihilation rate is expected to decline sharply as $X^-$ particles recombine with protons
and helium ions, and the Sommerfeld-Sakharov enhancement is reduced. Similarly, the annihilation rate is likely to change
significantly around $z\sim 1000$, when $X^+$ and $X^-\alpha$ particles recombine with electrons.  The present CHAMPs annihilation
rate would be very sensitive to the fractions of $X^-$ particles bound to different baryons.

\section{Summary}
We have found that CHAMPs with mass in the range between $100(q_X/e)^2 \lsim
m_X\lsim 10^8(q_X/e)$ TeV would be depleted from the galactic disk, which can
explain their previous non-detection. Though hard to detect, CHAMPs would make
a strong impact on the observable universe. In \S 4.2, we have shown that
through their interactions with magnetic fields, CHAMPs can affect the
structure and dynamics of baryonic matter. Conversely, baryons would make
a strong impact on the dark matter, in particular, in low-mass galaxies where
they would be able to flatten dark-matter cusps. CHAMPs with  unit charge and
mass of order a few hundred TeV seem especially attractive. Such particles
may also be able to explain the underabundance of dwarf galaxies, the absence
of cooling flows in the cores of galaxy clusters \cite{CN}, the present value
of $\Omega_M$ \cite{unitar}, the origin of  supermassive black holes, and the
reionization history of the Universe.

It has been proposed that some of the problems of $\Lambda$CDM model
may be solved by self-interacting dark matter (SIDM) \cite{SS}.
However, since CHAMPs interact with each other mainly through the intermediate agency
of magnetic fields rather than by direct scattering, they escape several problems
associated with the SIDM, such as predictions of too spherical halo shapes
\cite{Mir}, and unobserved evaporation of galactic halos inside galaxy clusters \cite{GO}.

\clearpage

\end{document}